\documentclass[12pt]{article}%
\usepackage{amsmath,latexsym}
\usepackage{graphicx}
\usepackage{amsmath}
\usepackage{amsfonts}
\usepackage{amssymb}%
\begin{document}

\title{{Supporting traversable wormholes: the case
   for noncommutative geometry}}
   \author{
Peter K. F. Kuhfittig*\\  \footnote{kuhfitti@msoe.edu}
 \small Department of Mathematics, Milwaukee School of
Engineering,\\
\small Milwaukee, Wisconsin 53202-3109, USA}

\date{}
 \maketitle

\begin{abstract}\noindent
While wormholes may be just as good a prediction
of Einstein's theory as black holes, they are
subject to severe restrictions from quantum field
theory.  In particular, a wormhole can only be
held open by violating the null energy condition,
calling for the existence of ``exotic matter."   
An equally serious problem
is the enormous radial tension at the throat of a
typical Morris-Thorne wormhole unless the wormhole
has an extremely large throat size.  It has been
proposed that noncommutative geometry, an offshoot
of string theory, may be the proper tool for
addressing these issues.  The purpose of this paper
is two-fold: (1) to make a stronger and more detailed 
case for this proposal and (2) to obtain a complete 
wormhole solution from the given conditions.
\end{abstract}

\section{Introduction}\label{E:introduction}

Both wormholes and black holes are predictions
of Einstein's theory, but they are subject to
severe restrictions from quantum field theory.
These include a violation of the null energy
condition (NEC), calling for the need for
``exotic matter."  Another serious problem is
the enormous radial tension at the throat.  To
address these issues, a number of modified
gravitational theories have been called upon.
For example, it was proposed by Lobo and
Oliveira \cite{LO09} that in $f(R)$ modified
gravity, the wormhole throat could be lined
with ordinary matter, while the unavoidable
violation of the NEC can be attributed to the
higher-order curvature terms.  Another
possibility is to hypothesize the existence
of an extra spatial dimension \cite{pK23a}.

It is proposed in this paper that noncommutative
geometry in the form presented by Nicolini et al.
\cite{NSS06} may be the best approach since it
is able to address each of these issues, as we
will see.

\section{Traversable wormholes}
The term ``wormhole" was coined by John
Archibald Wheeler in 1957, but the basic idea
can be traced to the Schwarzschild solution
and hence to black holes, which could be
viewed as nontraversable wormholes.  More
recently, the subject of entanglement has
led to the resurrection of a special type
of wormhole, the Einstein-Rosen bridge, to
explain this phenomenon \cite{MS13}.

The first detailed study of macroscopic
traversable wormholes is due to Morris and
Thorne \cite{MT88}, who proposed the
following static and spherically symmetric
line element for a wormhole spacetime:
\begin{equation}\label{E:line1}
ds^{2}=-e^{2\Phi(r)}dt^{2}+e^{2\alpha(r)}dr^2
+r^{2}(d\theta^{2}+\text{sin}^{2}\theta\,
d\phi^{2}),
\end{equation}
where
\begin{equation}
   e^{2\alpha(r)}=\frac{1}{1-\frac{b(r)}{r}}.
\end{equation}
(We are using units in which $c=G=1$.)  Morris
and Thorne also introduced the following
terminology: $\Phi=\Phi(r)$ is called the
\emph{redshift function}, which must be finite
everywhere to prevent the occurrence of an
event horizon (usually associated with black
holes). The function $b=b(r)$ is called the
\emph{shape function} since it determines the
spatial shape of the wormhole whenever it is
depicted in an embedding diagram \cite {MT88}.
The spherical surface $r=r_0$ is called the
\emph{throat} of the wormhole.  For what came
to be called a Morris-Thorne wormhole, some
additional requirements must be met: the
shape function has to satisfy the following
conditions: $b(r_0)=r_0$, $b(r)<r$ for $r>r_0$,
and $b'(r_0)<1$, called the \emph{flare-out
condition} in Ref. \cite{MT88}.  In classical
general relativity, this critical condition can
only be met by violating the null energy condition
(NEC), which states that for the energy-momentum
tensor $T_{\alpha\beta}$
\begin{equation}
   T_{\alpha\beta}k^{\alpha}k^{\beta}\ge 0\,\,
   \text{for all null vectors}\,\, k^{\alpha}.
\end{equation}
Matter that violates the NEC is called ``exotic"
in Ref. \cite{MT88}.  (The term was borrowed from
quantum mechanics.)  Exotic matter is going to
receive a lot of our attention.  The main reason 
is that for the radial outgoing null vector 
$(1,1,0,0)$, the violation becomes
\begin{equation}\label{E:violation}
   T_{\alpha\beta}k^{\alpha}k^{\beta}=\rho+
      p_r<0.
\end{equation}
Here $T^t_{\phantom{tt}t}=-\rho(r)$
is the energy density, $T^r_{\phantom{rr}r}=
p_r(r)$ is the radial pressure, and
$T^\theta_{\phantom{\theta\theta}\theta}=
T^\phi_{\phantom{\phi\phi}\phi}=p_t(r)$
is the lateral (transverse) pressure.  It
turns out, however, that Condition
(\ref{E:violation}) is not just
problematical, many researchers consider
this condition to be completely unphysical.

Our final requirement is \emph{asymptotic
flatness:}
\begin{equation}
   \text{lim}_{r\rightarrow\infty}\Phi(r)=0
   \quad \text{and} \quad
   \text{lim}_{r\rightarrow\infty}
   \frac{b(r)}{r}=0.
\end{equation}

We will finish this section by listing
the Einstein field equations, referring to
line element (\ref{E:line1}):

\begin{equation}\label{E:E1}
  \rho(r)=\frac{b'}{8\pi r^2},
\end{equation}
\begin{equation}\label{E:E2}
   p_r(r)=\frac{1}{8\pi}\left[-\frac{b}{r^3}+
   2\left(1-\frac{b}{r}\right)\frac{\Phi'}{r}
   \right],
\end{equation}
and
\begin{equation}\label{E:E3}
   p_t(r)=\frac{1}{8\pi}\left(1-\frac{b}{r}\right)
   \left[\Phi''-\frac{b'r-b}{2r(r-b)}\Phi'
   +(\Phi')^2+\frac{\Phi'}{r}-
   \frac{b'r-b}{2r^2(r-b)}\right].
\end{equation}

\section{The redshift function}
   \label{S:redshift}
Before continuing, we need to return to
Ref. \cite{MT88} to discuss the radial
tension at the throat.  We start by
recalling that the radial tension $\tau(r)$
is the negative of the radial pressure
$p_r(r)$.  According to Ref. \cite{MT88},
the Einstein field equations can be
rearranged to yield $\tau(r)$.  Temporarily 
reintroducing $c$ and $G$,  $\tau(r)$ is 
given by
\begin{equation}
   \tau(r)=\frac{b(r)/r-[r-b(r)]\Phi'(r)}
   {8\pi Gc^{-4}r^2}.
\end{equation}
The radial tension at the throat then becomes
\begin{equation}\label{E:tension}
  \tau(r_0)=\frac{1}{8\pi Gc^{-4}r_0^2}\approx
   5\times 10^{41}\frac{\text{dyn}}{\text{cm}^2}
   \left(\frac{10\,\text{m}}{r_0}\right)^2.
\end{equation}
As pointed out in Ref. \cite{MT88}, for
$r_0=3$ km, $\tau(r)$ has the same magnitude
as the pressure at the center of a massive
neutron star.  (For further discussion of
this problem, see Ref. \cite{pK22a}.)
Eq. (\ref{E:tension}) shows that for
$\tau(r)$ to have a small value, $r=r_0$
has to be extremely large.  Otherwise,
a wormhole is likely to be a compact stellar
object \cite{pK22b}.  (We will consider
another possibility in Sec. \ref{S:radial}.)

We can make our wormhole solution physically
acceptable by using a form of the redshift
function proposed by Lake \cite{kL03}:
\begin{equation}\label{E:Lake}
   2\Phi(r)=n\,\text{ln}(1+Ar^2),\quad n\ge 1,
\end{equation}
where $A$ is a positive constant.  According
to Ref. \cite{kL03}, we are dealing here with
a class of monotone increasing functions that
generate all regular static spherically
symmetric perfect-fluid solutions of the
Einstein field equations.  The reason that
Eq. (\ref{E:Lake}) is of particular interest
to us is  that it proved to be extremely
useful in the study of compact stellar
objects \cite{MDRK, MGRD}.  A more general
form is
\begin{equation}\label{E:Lake2}
   e^{2\Phi(r)}=B(1+Ar^2)^n,\quad A,B>0,
\end{equation}
where $A$ is still a free parameter, but
$B$ can be determined from the junction
conditions that allow the wormhole
spacetime to be joined to the exterior
Schwarzschild solution.  (See Ref.
\cite{pK22c} 
for details.)

\section{Noncommutative geometry}
      \label{S:noncommutative}
In this section we turn our attention to 
noncommutative geometry and its 
consequences.  Often presented as an 
offshoot of string theory, it assumes 
that point-like particles are replaced 
by smeared objects, an assumption that 
is consistent with the Heisenberg 
uncertainty principle.  The goal is 
to eliminate the divergences that 
normally occur in general relativity 
\cite{NSS06, SS03, NS10}.  According to 
Ref. \cite{NSS06}, this goal can be met 
by assuming that spacetime can be encoded
in the commutator $[\textbf{x}^{\mu},\textbf{x}^{\nu}]
=i\theta^{\mu\nu}$, where $\theta^{\mu\nu}$ is
an antisymmetric matrix that determines the
fundamental cell discretization of spacetime
in the same way that Planck's constant $\hbar$
discretizes phase space.  Fortunately, the 
smearing can be modeled by using a so-called 
Lorentzian distribution of minimal length 
$\sqrt{\beta}$ instead of the Dirac delta
function \cite{NM08, LL12}.  More precisely,
the energy density of a static and spherically
symmetric and particle-like gravitational
source is taken to be 
\begin{equation}\label{E:density}
   \rho(r)=\frac{m\sqrt{\beta}}{\pi^2
   (r^2+\beta)^2}.
\end{equation} 
According to this formulation, the 
gravitational source causes the mass $m$ to 
be diffused throughout the region of linear 
dimension $\sqrt{\beta}$ due to the 
uncertainty.

To make use of $\rho(r)$ in Eq. 
(\ref{E:density}), we need to recall the 
arguments proposed in Ref. \cite{NSS06}, 
starting with the critical assertion that 
the noncommutative effects can be 
implemented in the Einstein field 
equations 
$G_{\mu\nu}=\frac{8\pi G}{c^4}T_{\mu\nu}$
by modifying only the energy momentum
tensor $T_{\mu\nu}$, while leaving the
Einstein tensor $G_{\mu\nu}$ intact.  
The reasons given in Ref. \cite{NSS06} 
are complex: a metric field is a geometric 
structure defined over an underlying 
manifold whose strength is measured by 
its curvature, but the curvature, in turn, 
is nothing more than the response to the 
pressence of a mass-energy distribution.  
Furthermore, the noncommutativity is an 
\emph{intrinsic} property of spacetime 
rather than a superimposed geometric 
structure.  These observations justify 
the use of the above $\rho(r)$ in 
Eq. (\ref{E:E1}), restated here for 
convenience,
\begin{equation}\label{E:Einstein}
   \rho(r)=\frac{b'(r)}{8\pi r^2},
\end{equation}  
since the right-hand side is the 
$G_{tt}$ component of the Einstein 
tensor.  Moreover, according to Ref. 
\cite{NSS06}, the relationship between 
the radial pressure and the energy 
density is given by $p_r=-\rho$.

Before continuing we need to reexamine 
an earlier issue, the possible violation 
of the NEC at $r=r_0$.  Observe that since 
$p_r=-\rho$,   
\begin{equation}\label{E:violation2}
   T_{\alpha\beta}k^{\alpha}k^{\beta}=\rho+p_r
   =\frac{m\sqrt{\beta}}{\pi^2(r_0^2+\beta)^2}
   -\frac{1}{8\pi}\frac{b(r_0)}{r_0^3}<0
\end{equation}
since $\sqrt{\beta}\ll 1$.  Having justified 
the use of $\rho(r)$ from noncommutatve 
geometry, Inequality (\ref{E:violation2}) 
suggests that we have a violation of the 
NEC, \emph{at least locally}.  To move to 
a macroscopic scale, we will need the 
concept of \emph{emergence}, discussed in 
the next section.

\section{Emergence and its consequences}
     \label{S:emergence}
 The concept of emergence, which had its 
 roots in antiquity, is discussed in 
 Ref. \cite{pK23b} in the context of 
 traversable wormholes.  It is argued 
 that noncommutative geometry in the form 
 discussed in the previous section is a 
 \emph{fundamental property} and that the 
 outcome, a macroscopic wormhole, is an 
 \emph{emergent property}.  The basic 
 idea is that an emergent phenomenon is 
 derived from some fundamental theory.  
 For example, life emerges from objects 
 that are completely lifeless, such as 
 atoms and molecules.  It must be 
 emphasized, however, that such a process 
 is not reversible: living organisms tell 
 us nothing about the particles in the 
 fundamental theory.  These are no longer 
 relevant in the resulting \emph{effective 
 model}.  The local violation in Inequality 
 (\ref{E:violation2}) can now be seen as a 
 fundamental property and the resulting 
 macroscopic scale as an emergent property. 
 In other words, in the resulting effective 
 model, the short-distance effects are 
 discarded since these are meaningful only 
 in the fundamental theory. 
 
 Discussions of emergence often emphasize 
 the surprising or unexpected nature of the 
 outcome.  Our results are no exceptions.  
 The rest of this section will therefore be 
 devoted to confirming that the throat size 
 can indeed be macroscopic.  Our first step 
 is to return to Eq. (\ref{E:Einstein}) to 
 determine the shape function:
 \begin{multline}\label{E:shape1}
   b(r)=r_0+\int^r_{r_0}8\pi(r')^2\rho(r')dr'\\
   =\frac{4m}{\pi}
  \left[\text{tan}^{-1}\frac{r}{\sqrt{\beta}}
  -\sqrt{\beta}\frac{r}{r^2+\beta}-
  \text{tan}^{-1}\frac{r_0}{\sqrt{\beta}}
  +\sqrt{\beta}\frac{r_0}{r_0^2
  +\beta}\right]+r_0\\
  =\frac{4m}{\pi}\frac{1}{r}
  \left[r\,\text{tan}^{-1}\frac{r}{\sqrt{\beta}}
  -\sqrt{\beta}\frac{r^2}{r^2+\beta}-
  r\,\text{tan}^{-1}\frac{r_0}{\sqrt{\beta}}
  +\sqrt{\beta}\frac{r_0r}{r_0^2
  +\beta}\right]+r_0.
\end{multline}
We can now follow Ref. \cite{pK23b}, which 
unexpectedly shows that $B=b/\sqrt{\beta}$
has the properties of a shape function.  
The reason is that $B$ can be readily 
expressed as a function of $r/\sqrt{\beta}$:
\begin{multline}\label{E:shape2}
   \frac{1}{\sqrt{\beta}}
   \,b(r)=
   B\left(\frac{r}{\sqrt{\beta}}\right)=\\
   \frac{4m}{\pi}\frac{1}{r}\left[\frac{r}{\sqrt{\beta}}
   \,\text{tan}^{-1}\frac{r}{\sqrt{\beta}}
   -\frac{\left(\frac{r}{\sqrt{\beta}}\right)^2}
   {\left(\frac{r}{\sqrt{\beta}}\right)^2+1}
  -\frac{r}{\sqrt{\beta}}\,
  \text{tan}^{-1}\frac{r_0}{\sqrt{\beta}}
  +\frac{r}{\sqrt{\beta}}
  \frac{\frac{r_0}{\sqrt{\beta}}}
  {\left(\frac{r_0}{\sqrt{\beta}}\right)^2+1}
  \right]+\frac{r_0}{\sqrt{\beta}}.
\end{multline}
 We now have
 \begin{equation}\label{E:throat}
   B\left(\frac{r_0}{\sqrt{\beta}}\right)
   =\frac{r_0}{\sqrt{\beta}},
\end{equation}
the analogue of $b(r_0)=r_0$.  It follows 
that the throat size is macroscopic, 
confirming that we are indeed dealing 
with an emergent property.  It is also clear 
from Eqs. (\ref{E:E1}) and (\ref{E:shape1}) 
that the mass $m=\frac{1}{2}b$ of the wormhole 
depends on the initial condition $b(r_0)=r_0$.    

Another issue that needs to be addressed 
is the enormous radial tension at the 
throat.  A seemingly insurmountable problem,
noncommutative geometry offers a solution.  
That is the topic of the next section.

\section{General problems with low-density 
   wormholes}\label{S:general}
   
The term ``low density" when referring to 
wormholes is merely a reference to the use 
of geometrized units.  For example, for 
nuclear matter, whose density is 
$10^{18}\text{kg}/\text{m}^3$, we obtain
\begin{equation}
   10^{18}\frac{\text{kg}}{\text{m}^3}
   \frac{G}{c^2}=7.41\times 10^{-2}
   \,\text{m}^{-2}.
\end{equation}
So from Eq. (\ref{E:Einstein}), 
\begin{equation}\label{E:flare}
   b'(r)=8\pi r^2\rho(r)<1,
\end{equation}            
even for much larger $\rho(r)$.  So 
the flare-out condition is satisfied 
for what is essentially an arbitrary 
$\rho(r)$.  So without the special 
features from noncommutative geometry, 
we cannot draw the same conclusions 
using only Eq. (\ref{E:flare}).  This 
can also be illustrated by means of a 
simple example, the zero-density case 
$\rho\equiv 0$, treated in Visser's 
book \cite{mV95}.  We get a valid 
wormhole solution only if we require 
the usual exotic matter.  The same 
comments apply to both dark matter and 
dark energy, each of which has an 
extremely low energy density.  (The 
dark-matter case also raises an obvious 
question: if exotic matter is needed 
anyway, then what is the role of dark 
matter in the first place?)

The dark-energy case needs some 
clarification, however: recall that 
given its (isotropic) equation of state 
$p=\omega\rho$, whenever $\omega<-1$ we 
are dealing with a special kind called 
``phantom energy."  Observe that for this 
case, $\rho+p=\rho+\omega\rho=
\rho(1+\omega)<0$.  Since the NEC has been 
violated, phantom dark energy could in 
principle support traversable wormholes.  
In summary, neither dark matter alone nor 
dark energy alone can support a 
Morris-Thorne wormhole: the former 
requires the existence of exotic matter 
and the latter the equation of state 
$p=\omega\rho$, $\omega<-1$.  For further 
discussion of this issue, see Ref. 
\cite{pK23c}.  

\section{The radial tension at the throat}
   \label{S:radial}
Our final topic is the enormous radial 
tension at the throat, mentioned earlier.  
According to Eq. (\ref{E:tension}), the 
magnitude of the radial tension is directly 
connected to the throat size; so low-tension 
wormholes could only exist on very large 
scales.  As noted after Eq. (\ref{E:tension}), 
for $r_0=3 \,\text{km}$, $\tau(r)$ would 
have the same magnitude as the pressure at 
the center of a massive neutron star.  
Attributing this outcome to exotic matter 
ignored the fact that exotic matter was 
introduced for a completely different reason, 
ensuring the violation of the NEC.  This 
takes us back to the subject of this paper.  
It is shown in Ref. \cite{pK20} that a 
noncommutative-geometry background can 
account for the high radial tension by 
starting with the observation that the 
smeared particles that comprise the throat 
surface will cause the surface itself to 
be smeared.  Its density $\rho_s$ is given 
by \cite{pK20}
\begin{equation}\label{E:surface1}
   \rho_s=\frac{\mu\sqrt{\beta}}
   {\pi^2[(r-r_0)^2+\beta]^2},
\end{equation}          
where $\mu$ is the mass of the surface.  
Since the tension $\tau$ is the negative 
of the pressure, the violation of the NEC, 
$p_r+\rho_s<0$, now becomes $\tau-\rho_s>0$ 
at the throat.  The condition $p_r=-\rho_s$ 
implies that we are right on the edge of 
violating the NEC, i.e., $\tau-\rho_s=0$. 
So from Eq. (\ref{E:surface1}), we have at 
$r=r_0$,
\begin{equation}\label{E:surface2}
   \rho_s=\frac{\mu}{\pi^2}\frac{1}
        {\beta^{3/2}}.
\end{equation}
Since the surface is smeared, we can 
only say that $r\approx r_0$; it therefore 
follows from Eq. (\ref{E:surface1}) that 
$\rho_s$ is reduced in value.  So instead 
of $\tau-\rho_s=0$, we actually have 
the desired
\begin{equation}
   \tau-\rho_s>0.
\end{equation}        
Furthermore, Eq. (\ref{E:surface2}) 
implies that $\rho_s$ and hence $\tau$ 
are extremely large, which was to be 
shown.

Returning to Eq. (\ref{E:tension}), we 
now see that for a throat size of 10 m, 
$\tau\approx 5\times 10^{41} \, 
\text{dyn}/\text{cm}^2$.  As noted in Ref. 
\cite{pK20}, if $\mu$ has the rather 
minute value of $10^{-10}$ g, then from 
Eq. (\ref{E:surface2}), we get 
\begin{equation}
  \tau=\rho_s c^2=\frac{\mu}{\pi^2}
  (\sqrt{\beta})^{-3}c^2=5\times 10^{41}\, 
  \frac{\text{dyn}}{\text{cm}^2}.
\end{equation}
To satisfy this relationship, the value 
of $\sqrt{\beta}=10^{-11}$ cm is sufficient.  
Since $\sqrt{\beta}$ may be much smaller, 
we can accommodate even larger values of 
$\tau$.

The unexpected nature of this result 
suggests that the high radial tension is 
another example of an emergent phenomenon.

\section{Conclusion}\label{S:Conclusion}

Both black holes and wormholes are 
predictions of Einstein's theory, but 
wormholes are subject to severe restrictions 
from quantum field theory.  In particular, 
a wormhole can only be held open by 
violating the null energy condition, calling 
for the existence of ``exotic matter."   
An equally serious problem is the enormous 
radial tension at the throat, unless the 
wormhole has an extremely large throat size.

While these problems may seem insurmountable, 
it has been suggested that noncommutative 
geometry, an offshoot of string theory, 
may be the proper tool for addressing these 
issues.  The purpose of this paper is to 
make a stronger and more detailed case for 
this assertion.

The discussion of noncommutative geometry 
starts in Sec. \ref{S:noncommutative}, 
following a general introduction to 
traversable wormholes.  The energy density 
of a static and spherically symmetric and 
particle-like gravitational source is given 
in Eq. (\ref{E:density}).  The implication 
is that the mass $m$ of a particle is 
diffused throughout the region of linear 
dimension $\sqrt{\beta}$ due to the 
uncertainty.  In Sec. \ref{S:radial}, 
the idea of a smeared particle is 
extended to that of a  smeared surface, 
the throat of a wormhole.  

The discussion in Sec. \ref{S:noncommutative} 
continues with the critical assertion that 
the noncommutative effects can be implemented 
in the Einstein field equations by modifying 
only the energy momentum tensor, while leaving 
the Einstein tensor intact.  The result is a 
geometric structure defined over an underlying 
manifold.  We conclude that noncommutativity 
is an \emph{intrinsic} property of spacetime 
rather than a superimposed geometric structure.
The outcome, Eq. (\ref{E:violation2}), can be 
viewed as a violation of the NEC \emph{at 
least locally}.  Moving to a macroscopic scale 
requires the concept of \emph{emergence}, 
discussed in Sec. \ref{S:emergence}.  Here it 
is emphasized that emergence often results in 
surprising and unexpected outcomes.  The rest 
of the section is therefore devoted to 
confirming this approach by showing explicitly 
that the throat size is indeed macroscopic 
thanks to the noncommutative-geometry 
background. 

Sec. \ref{S:general} addresses a very general 
problem regarding the energy density $\rho(r)$ 
stemming from the use of geometrized units: the 
flare-out condition tends to be satisfied for 
an arbitrary $\rho(r)$ since $b'(r)=
8\pi r^2\rho(r)<1$ due to the small value of 
$\rho(r)$ in geometrized units.  Without the 
special features from noncommutative geometry, 
we cannot safely conclude that $b'(r_0)<1$ 
for an arbitrary $\rho(r)$.  These include 
solutions based on a dark-matter or 
dark-energy background.  In other words, 
neither dark matter alone nor dark energy 
alone can support a Morris-Thorne wormhole: 
the former requires the existence of exotic 
matter and the latter the equation of state 
$p=\omega\rho$, $\omega<-1$.

Sec. \ref{S:radial} deals with another 
aspect of noncommutative geometry, the smeared 
throat surface mentioned earlier.  Using 
the expression for the energy density in 
Eq. (\ref{E:surface1}), it is shown that the 
noncommutative-geometry background can account 
for the enormous radial tension at the throat.  
The unexpected nature of this result once again
points to emergence.  

Our final topic is concerned with the very 
nature of our wormhole solution.  The original 
Morris-Thorne paper \cite{MT88} used a reverse 
strategy: first assign both the redshift and 
shape functions and then search the Universe 
for matter or fields that would produce the 
desired stress-energy tensor.  In this paper  
we obtained a complete wormhole solution by 
using the opposite approach.  As noted in Sec. 
\ref{S:redshift}, since a traversable 
wormhole may very well be a compact stellar 
object, we can use a form of the redshift 
function proposed by Lake \cite{kL03}, 
slightly generalized in Eq. (\ref{E:Lake2}).     
This form is not only physically acceptable, 
it is general enough to include wormholes 
that are compact stellar objects.  The 
shape function was determined from the 
noncommutative-geometry background.  The 
result is a complete wormhole solution 
determined from the given conditions.

\end{document}